\documentclass[sigconf]{acmart}
\usepackage{subcaption}
\usepackage{changepage}
\usepackage{multirow}
\usepackage{xcolor}
\usepackage{soul}
\usepackage{enumitem}

\usepackage{censor}

\definecolor{drk_blue}{HTML}{263761}
\definecolor{med_blue}{HTML}{4F71BE}
\definecolor{lt_blue}{HTML}{B7C6E4}
\definecolor{very_lt_blue}{HTML}{DBE3F1}
\definecolor{white}{HTML}{FFFFFF}

\AtBeginDocument{%
  \providecommand\BibTeX{{%
    \normalfont B\kern-0.5em{\scshape i\kern-0.25em b}\kern-0.8em\TeX}}}

\copyrightyear{2024}
\acmYear{2024}
\setcopyright{rightsretained}
\acmConference[ICSE-SEIS'24]{Software Engineering in Society}{April 14--20, 2024}{Lisbon, Portugal}
\acmBooktitle{Software Engineering in Society (ICSE-SEIS'24), April 14--20, 2024, Lisbon, Portugal}\acmDOI{10.1145/3639475.3640103}
\acmISBN{979-8-4007-0499-4/24/04}
\begin{document}




\title[Towards Fair and Equitable Software Low-Altitude Airspace Authorizations]{Towards Engineering Fair and Equitable Software Systems for Managing Low-Altitude Airspace Authorizations}

\author{Usman Gohar}
\affiliation{%
 \institution{Iowa State University}
 \city{Ames}
 \state{Iowa}
 \country{USA}}
 \email{ugohar@iastate.edu}

\author{Michael C. Hunter }
\affiliation{%
 \institution{Iowa State University}
  \streetaddress{30 Shuangqing Rd}
 \city{Ames}
 \state{Iowa}
 \country{USA}}
 \email{mchunter@iastate.edu}

\author{Agnieszka Marczak-Czajka}
\affiliation{%
  \institution{University of Notre Dame}
  \city{South Bend}
  \state{Indiana}
  \country{USA}
  \postcode{78229}}
\email{amarczak@nd.edu}

\author{Robyn R. Lutz}
\affiliation{%
 \institution{Iowa State University}
  \streetaddress{1 Th{\o}rv{\"a}ld Circle}
 \city{Ames}
 \state{Iowa}
 \country{USA}}
\email{rlutz@iastate.edu}

\author{Myra B. Cohen}
\affiliation{%
 \institution{Iowa State University}
 \city{Ames}
 \state{Iowa}
 \country{USA}}
\email{mcohen@iastate.edu}

\author{Jane Cleland-Huang}
\affiliation{%
  \institution{University of Notre Dame}
 \city{South Bend}
 \state{Indiana}
 \country{USA}}
\email{janehuang@nd.edu}

\renewcommand{\shortauthors}{Usman Gohar, Michael C. Hunter, Agnieszka Marczak-Czajka, et al.}

\ccsdesc[500]{Software and its engineering~Requirements analysis}

\keywords{Drones, sUAS, fairness, machine learning, software engineering}

\begin{abstract}
\label{sec:abstract}

Small Unmanned Aircraft Systems (sUAS) have gained widespread adoption across a diverse range of applications. This has introduced operational complexities within shared airspaces and an increase in reported incidents, raising safety concerns. In response, the U.S. Federal Aviation Administration (FAA) is developing a UAS Traffic Management (UTM) system to control access to airspace based on an sUAS's predicted ability to safely complete its mission. However, a fully automated system capable of swiftly approving or denying flight requests can be prone to bias and must consider safety, transparency, and fairness to diverse stakeholders. In this paper, we present an initial study that explores stakeholders' perspectives on factors that should be considered in an automated system. Results indicate flight characteristics and environmental conditions were perceived as most important but pilot and drone capabilities should also be considered.
Further, several respondents indicated an aversion to any AI-supported automation, highlighting the need for full transparency in automated decision-making. Results provide a societal perspective on the challenges of automating UTM flight authorization decisions and help frame the ongoing design of a solution acceptable to the broader sUAS community.

\end{abstract}

\maketitle

\section*{General abstract}

Small Unmanned Aircraft Systems (sUAS), also called drones, are becoming increasingly popular due to their usefulness in applications such as emergency response, remote sensing, agriculture, and package delivery, as well as with hobbyists. With the rise of sUAS operations, there has been an inevitable corresponding rise in the number of incidents, such as crashes or airspace infringements with other aircraft or in no-fly zones. 
For example, a package delivery drone should not interfere with airplanes or first-responder drones. Developing, validating, and deploying an automated system to approve or reject such sUAS flight requests is a complex, multi-faceted 
process that ultimately must balance competing concerns, including safety, transparency, and fairness to diverse stakeholders. This paper reports results from an initial survey we conducted with remote sUAS pilots, emergency responders, and members of the general public. The survey elicited their opinions about factors they perceive to be important in an automated authorization decision process.  
Survey results show that flight characteristics and environmental conditions were perceived as most important and that pilot and drone capabilities should also be considered. These results open up communication and help inform the  
human aspects inherent in the design and acceptance of such an automated system.

\section{Introduction}
\label{sec:introduction}

The deployment of small Unmanned Aircraft Systems (sUAS) within the US National Airspace has seen dramatic growth with applications such as remote sensing, package delivery, and emergency response \cite{8682048,Erdelj2017HelpFT,FAA}. These operations are conducted in airspace shared with other sUASs' and constrained by no-fly zones such as airports, national parks, and schools. The rapid escalation in sUAS numbers has been accompanied by a corresponding surge in reported incidents, often attributed to issues such as hardware or software malfunctions, human errors, including reckless disregard for rules and regulations, or external factors such as radio interference and adverse weather conditions \cite{dronecenter.bard.edu_2019,FAA}. Consequently, sUAS operators must seek permission to fly in all controlled airspace. 
For example, in the USA, Remote Pilots In Command (RPICs) must currently request flight permission through the Low Altitude Authorization and Notification Capability (LAANC), which grants access to airspace below 400 feet AGL (above ground level), provides awareness for where RPICs can and cannot fly, and provides visibility to air traffic controllers into where sUAS are currently operating \cite{LAANC}. 
The current system does not take into consideration specific flight details, environmental factors, drone characteristics, or pilot competencies.

To this end, a unified ecosystem called the UAS Traffic Management System (UTM) is being developed by the FAA, with research support from NASA, to coordinate large numbers of sUAS in shared low-altitude airspace \cite{UTM}. The UTM system relies on the digital exchange of planned flight information for each RPIC. An sUAS's capability to successfully and safely execute a mission is influenced by factors such as its inherent features (e.g., aircraft weight, size, and onboard sensors, etc.), its suitability for flight, operating environment (weather conditions, population density, the complexity of airspace, etc.) the planned flight characteristics, and the human operator (track record, license,  certification per FAA regulation - referred to as Part 107 in the USA, and skills, etc.) \cite{policyorder, UTMImplementationPlan}. While currently still under development, the UTM would mandate that operators submit a Performance Authorization Request (PAR) detailing how the sUAS's ground assets, services, personnel, and maintenance protocols will ensure safe operation for the flight duration. 
The software system to be developed for the UTM  is safety-critical in that its decisions can contribute to the safety of an airspace or can compromise its safety \cite{oLeve11}. Moreover, some sUAS operations undertaken in an airspace are themselves safety-critical, such as rescue operations, and could be jeopardized or delayed by an injudicious or unintended decision by the software.

Evaluating the PAR is currently a manual, labor-intensive process and, therefore, lacks scalability and is prone to human error. Without automation, humans cannot make decisions fast enough to handle the anticipated load of sUAS requesting entry to populated airspaces \cite{Reichmann_2021}. More importantly, many flight requests are currently granted without considering the historical performance of the sUAS, its RPIC, or its supporting organizational infrastructure, relying solely on self-declared compliance statements \cite{LAANC}.

These challenges might be addressed by developing a fully automated, 
flight authorization system that promptly grants or denies permissions for all but the most complex flight requests \cite{9348605}. At first, the system may be heuristic or rule-based; however, as data is gathered incrementally over time, the system likely will need to incorporate ML for scalability.  While the adoption of an ML-based approach appears to offer clear advantages in terms of supporting scalability, the opaque \cite{10.1145/3338906.3338937} nature of such systems also raises valid societal concerns regarding algorithmic fairness and transparency of automated flight authorizations \cite{angwin2022machine,10.1145/3236024.3264838}. 

Measuring and mitigating discrimination in ML/AI systems has been studied extensively by the software engineering and ML/AI communities \cite{mehrabi2021survey,gohar2023towards}. Mostly, these studies have been facilitated by the availability of large amounts of relevant data that have been used to train these ML systems and subsequently evaluate them for the presence of bias \cite{gohar2023survey}. Given that UTM has only been tested in controlled pilot tests so far and that the LAANC system does not collect detailed data about flight requests, no extensive datasets are currently available for training or automatically evaluating  PARs. 

This creates both a challenge and an opportunity for engineering such a system. The challenge exists because, unlike many other ML/AI automated decision-making systems, such as those related to determining prison sentences or mortgages  \cite{chouldechova2017fair,barocas2017fairness}, there is no existing data, biased or not, for training an initial system. The opportunity exists, therefore, however difficult it may be, to build a greenfield system that does not build upon decades of potentially biased human decisions but incorporates fairness and transparency considerations now while UTM architectures are still being developed. The perception of unfairness in decisions, particularly when they seem to benefit specific community segments at the expense of others, together with the transparency of these decisions, significantly affects their acceptance within the broader community \cite{matsuyama2021determinants,Gross2007CommunityPO}. As a result, we want to understand which factors are important to stakeholders of the UTM.


As a starting point in this process, we conducted an initial exploratory study aimed at eliciting concerns and issues that could provide insights into what a fair and equitable automated flight authorization system that decides whether to authorize an sUAS to access controlled airspace might look like.  
Specifically, we conducted a survey study to identify i) the factors deemed important by stakeholders in automated flight authorization decisions; and ii) the challenges requiring resolution for wider community acceptance. This work represents the first step 
of involving the sUAS community in discovering the requirements for a fair and equitable software system for managing low-altitude airspace authorization.

The rest of the paper is organized as follows. Section \ref{sec:background} describes the UTM ecosystem, a high-level overview of the automated flight authorization system, and the motivation for our work. In Section \ref{sec:methodology}, we discuss our study methodology, including survey design, data collection, and data analysis. In
Section \ref{sec:results}, we present the results of our study, followed by a discussion of lessons learned and future work in Section \ref{sec:discussion}. We describe threats to validity in Section \ref{sec:threats}, related works in Section \ref{sec:related}, and present the conclusion in Section \ref{sec:conclusion}.

\section{Background and Motivation}
\label{sec:background}

In this section, we discuss the concept of UTM and provide an overview of an automated sUAS flight authorization system. Additionally, we present a real-world motivating example to illustrate the need for our study.
\begin{figure*}[h]
    \centering
    \includegraphics[]{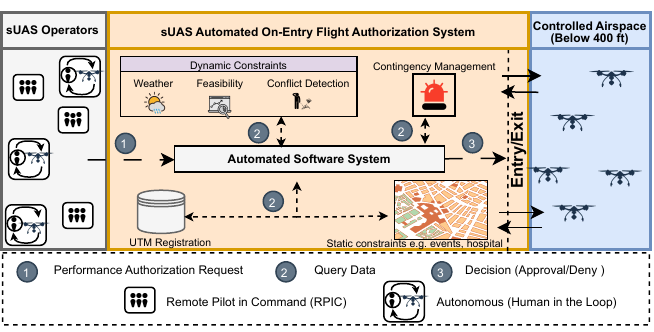}
    \caption{Overview of an automated software system for managing low-altitude airspace authorizations. In Step 1, the operator submits a PAR. Step 2 analyzes various factors about the flight, pilot, mission, and environment and makes a decision. Step 3 grants or denies permission based on the analysis from Step 2. }
    \label{SADE}
\end{figure*}

\subsection{UAS Traffic Management (UTM)}
sUAS typically fly at low altitudes (i.e., below 400 feet in the USA) with an RPIC as the operator \cite{10.1145/3468264.3468534}. The need for a dedicated UTM system is driven by three primary challenges. Firstly, the anticipated expansion in sUAS activities may soon rival, if not surpass, the current levels of manned air traffic \cite{kopardekar2019unmanned}. Recent research suggests that by 2035, the congestion levels in this particular airspace will escalate significantly, potentially reaching up to 65,000 operations per hour with a total market value of 10 billion euros per year \cite{sesar2016european}. Secondly, sUAS are expected to use air space differently from manned aircraft \cite{hassanalian2017classifications}. In addition to operating in low-altitude airspace within densely populated regions, the diversity of sUAS flight operations, ranging from package delivery and emergency response to traffic monitoring, introduces a wide array of complex and diverse operational scenarios (e.g., Beyond Visual Line of Sight or BVLOS) \cite{Erdelj2017HelpFT}. Ensuring the safe integration of new entrants alongside existing users is critical \cite{zhang2020collision}. Finally, governments and regulatory bodies have concluded that current Air Traffic Management cannot support the expected sUAS traffic volume and the challenges that integration of sUAS operations would bring \cite{jiang2016unmanned,ICAO}. Consequently, the FAA and NASA introduced the UTM concept to manage the increasing number of drone operations at low altitudes \cite{UTM}. The UTM is an ecosystem made up of multiple services (e.g., weather, emergency management, etc.)  and systems that enable the safe operations of sUAS in shared and isolated airspace. 

Recently, the FAA published a list of essential factors that must be assessed as a minimum requirement for conducting a risk assessment \cite{policyorder}. To be granted access to UTM-controlled airspace, an sUAS operator must submit a PAR that documents the sUAS's capabilities with respect to those expected by the FAA. A team of safety analysts uses this documentation to determine whether the planned mission is likely to be completed successfully. With projected operations expected to grow manyfold, automated flight authorization software systems could play a significant role.

We provide a high-level overview of a future automated flight authorization system in Figure \ref{SADE}. The first step involves an sUAS operator submitting the PAR (step 1) to the automated flight authorization software system. The system takes into account different information (step 2) about the flight and the pilot, including, but not limited to, the mission details and dynamic conditions such as current air traffic and weather. Based on this, the automated system either grants or denies permission for the flight (step 3). 

In contrast to ML/AI systems that are initially trained on existing data from prior human decisions, the UTM domain lacks a significant body of historical data that could be used to train an ML/AI software system or influence how the automated flight authorization system should be built. Therefore, the study reported in this paper explores the stakeholders' perspectives on these factors and their concerns regarding such an automated system.

\subsection{Motivating example}

\begin{figure}[]
    \centering
       \begin{subfigure}[t]{1\linewidth}
            \centering            
            \includegraphics[width=\linewidth]{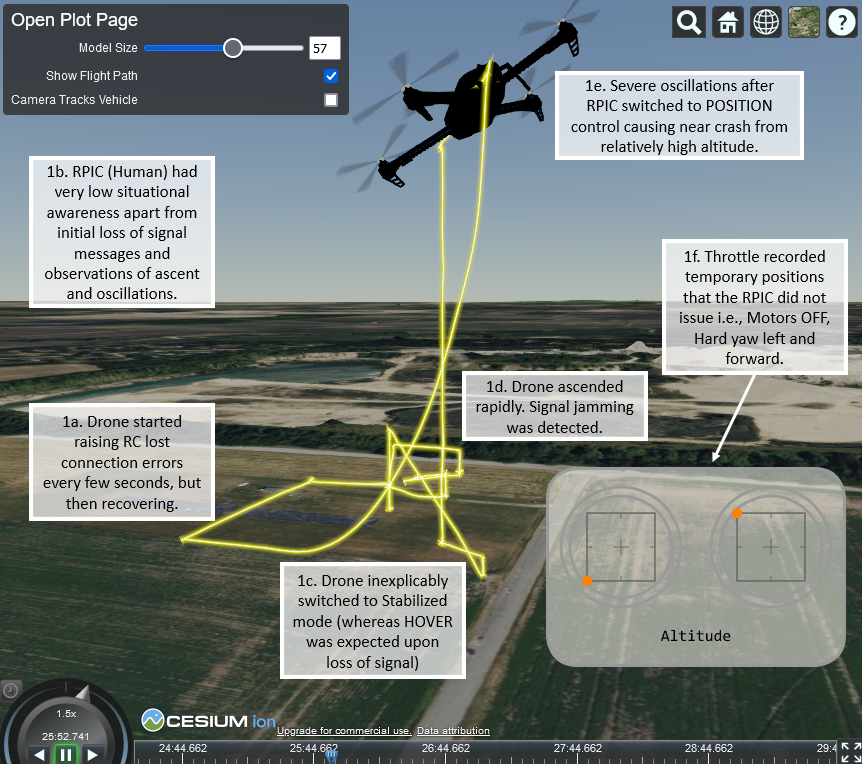} 
            \caption{Flight 1: The mode inexplicably switched to \textsc{stabilized}, causing the sUAS to ascend rapidly to 100 feet. Interference then caused erratic flight behavior and severe oscillations. The RPIC landed the sUAS manually, and it was grounded for repairs.           
            \vspace{8pt}}
           \label{fig:flight1}
            \end{subfigure}

       \begin{subfigure}[t]{1\linewidth}
            \centering            
            \includegraphics[width=\linewidth]{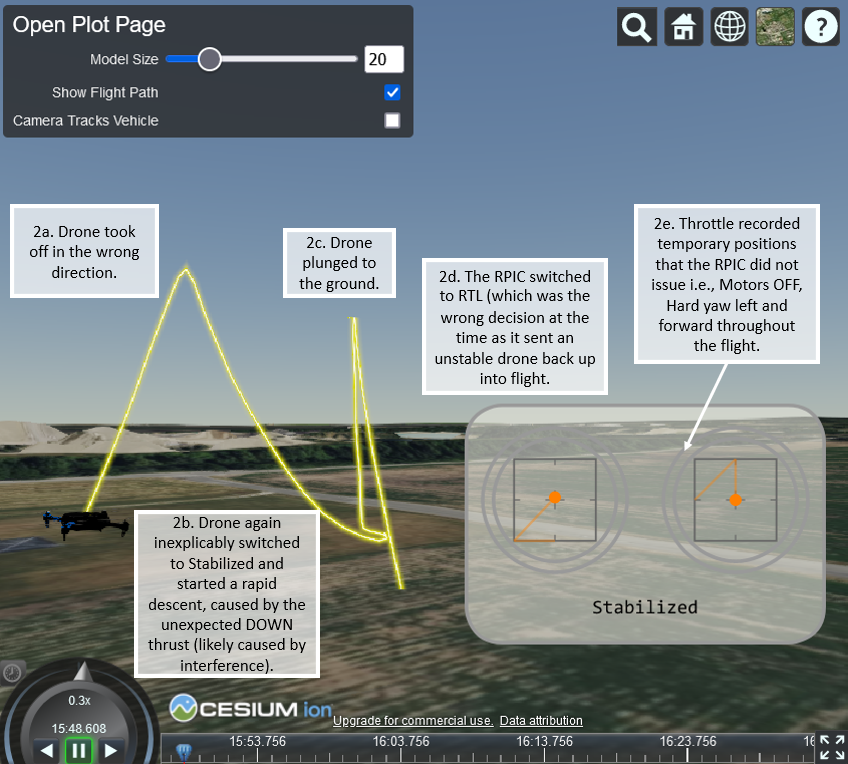}
            \caption{Flight 2: A few days later, another RPIC  flew the \textit{same} sUAS without performing maintenance. Similar interference and flight issues were observed, resulting in a high-altitude crash.}
            \label{fig:flight2}
            \end{subfigure}
            \caption{Flight replays of two physical sUAS flights.} 
             \vspace{-1mm}
        \label{fig:flightexample}
 \end{figure}

The proliferation of drone operations has underscored the need for a UTM ecosystem to manage escalating numbers of sUAS. Without an automated flight authorization system, manual reviews of PARs are insufficient to accommodate the potential workloads of such a system \cite{Reichmann_2021a}. The current rule-based system (LAANC) \cite{wallace2018evaluating} primarily focuses on information pertaining to air traffic, such as Temporary Flight Restrictions and Notices to Airmen. This overlooks vital safety assessment factors outlined by the FAA for the UTM concept \cite{policyorder,UTMImplementationPlan}. No works have evaluated the key factors to be considered when authorizing sUAS flights and how each factor influences the decision from the stakeholders' perspectives.
For example, consider the sUAS flights depicted in Figure \ref{fig:flightexample}. Both cases are replays of flights flown by the same physical hexacopter under the supervision of an RPIC.  In the first flight (Figure \ref{fig:flight1}), the sUAS experienced intermittent interference shortly after takeoff, which the supervising RPIC ignored. A few minutes later, the sUAS inexplicably switched from its autonomous `offboard' mode to `stabilized' mode and rapidly ascended to 377 meters. The supervising RPIC assumed control of the sUAS, at which point the sUAS experienced severe oscillations, almost resulting in a high-altitude crash. The sUAS was grounded for repairs. 

In the second flight (Figure \ref{fig:flight2}), a different RPIC flew the same sUAS without conducting any repairs, recklessly ignoring the fact that it had been grounded. Similar interference occurred, again causing strange flight behavior, resulting in a crash. If an automated flight authorization request system had been deployed, recent sUAS history, pilot history, and lack of maintenance could all have factored into a flight authorization decision. Furthermore, the second RPIC could have been required to provide documentation explaining how the previous problem had been addressed before the flight was authorized. This example highlights the variety of issues that could factor into a decision -- including ones that go beyond the current flight characteristics. For example, how much should a pilot's or drone's prior incident history impact the system's decision? How important is the interplay of pilot experience and weather conditions? 

Building an automated flight authorization system that makes UTM on-entry decisions is relatively easy, but ensuring that the decisions it makes respect the needs and desires of a broad set of stakeholders is much more challenging. Therefore, this study aims to explore stakeholders' perspectives toward a fair and equitable sUAS automated flight authorization system.

\section{Methodology}
\label{sec:methodology}

We conducted a preliminary exploration study to answer the following research questions:

\begin{itemize}[leftmargin=*]
    \item \textbf{RQ1}:\textit{What are the important factors that 
    stakeholders report are needed to make fair, equitable, and safe decisions?}
    \item \textbf{RQ2:} \textit{What are the stakeholder concerns related to the automated flight authorization system?}
\end{itemize}

We used a mixed-method research approach \cite{creswell1999mixed}, combining qualitative and quantitative analysis to answer our research questions. Participation in the study was voluntary. The remainder of this section outlines our survey methodology, including design, data collection, and analysis techniques. 

\subsection{Survey Design}

We drew inspiration from the work of \citeauthor{10.1145/3236024.3264833} \cite{10.1145/3236024.3264833} and used vignettes to understand the perspectives of the community on an automated flight authorization system. A vignette is a concise yet descriptive illustration of individuals or scenarios commonly employed to explore and understand human behavior and decision-making \cite{atzmuller2010experimental}. Vignettes provide a standardized and simplified way to present complex scenarios and minimize potential response bias that might arise from explicitly stating the manipulated factor \cite{atzmuller2010experimental,alexander1978use}. We used vignettes to carefully design and present multiple flight authorization request scenarios to the participants. We chose vignettes rather than interviews to increase participation and coverage of the study.

\begin{table*}[t]
\caption{Constructs, levels, and attributes identified prior to the study.}
\label{table:contructs}
\setlength\tabcolsep{1pt}
\begin{tabular}{c|c|c}
\hline
\textbf{Constructs} & \textbf{Levels} & \textbf{Attributes} \\
\hline
Pilot      & Novice, Average, Experienced & Simulated/Actual Flight Hours, Adverse Condition Flight Experience, Pilot Incidents \\
Drone      & Poor, Average, Good          & Flight History, Physical History, Make/Model                                        \\
Conditions & Poor, Average, Good          & Air Traffic, Human Traffic, Weather Conditions, Physical Area                       \\
Mission    & Easy, Average, Difficult     & Purpose, Complexity Level, Characteristics, Simulations    \\       \hline   
\end{tabular}
\end{table*}

\begin{figure*}[]
    \centering
    \includegraphics[scale = 0.38]{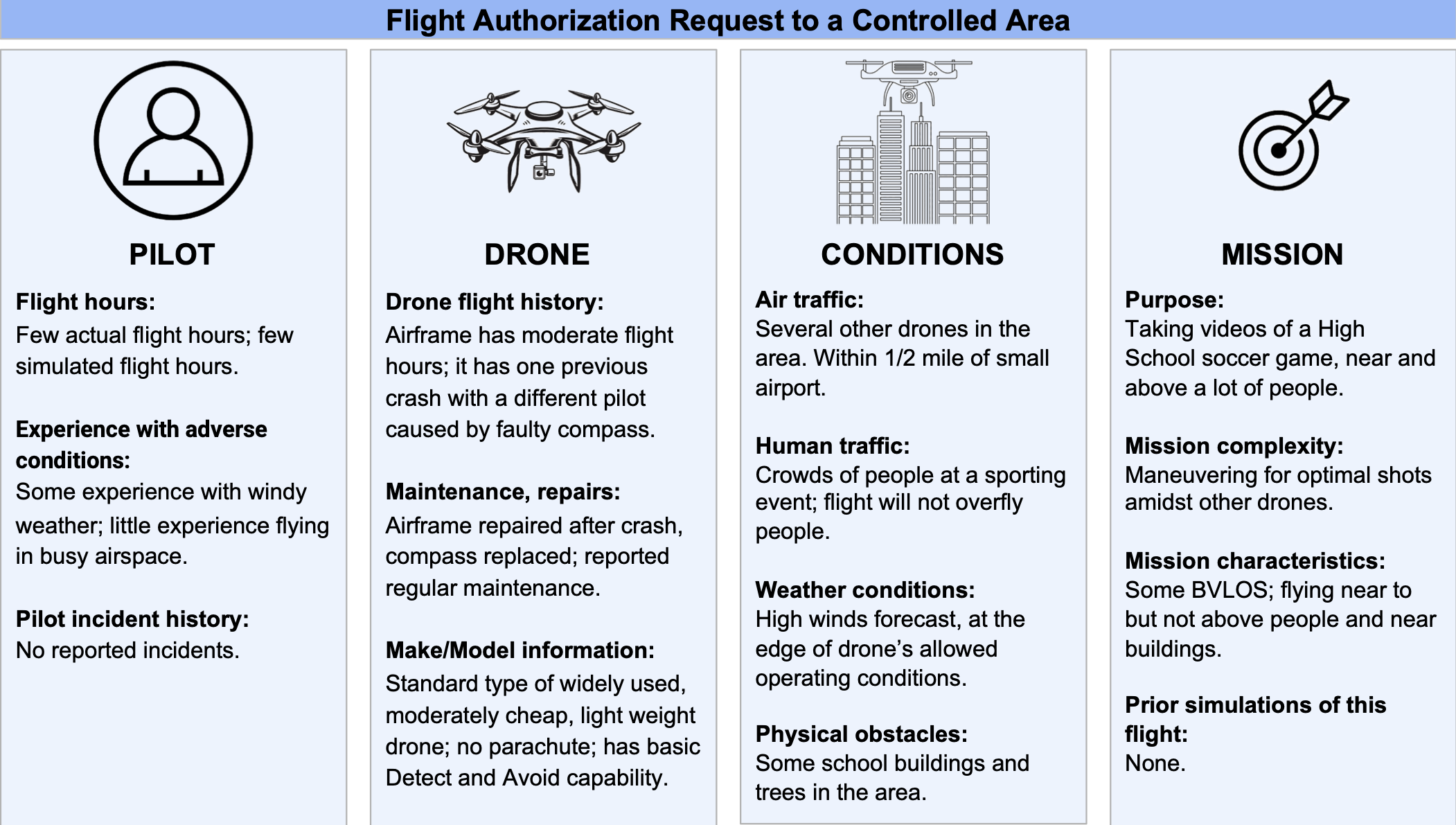}
    \caption{Study participants were asked to approve or reject a flight authorization request presented in the form of a vignette. Each vignette provided a high-level summary rather than a detailed authorization request, allowing respondents to assess the importance of pilot, drone, environmental, and mission concepts rather than focusing on low-level details. }
    \label{fig:vignette}
\end{figure*}

\subsubsection{Vignette Design Stage}

During the initial design phase, we developed a pilot survey by drawing insights from various sources, including studies \cite{10.1145/3468264.3468534} and incident reports \cite{report1, report2, report3, GOVUK2023, Wikipedia2023, barrado2020u}, reviewing approved FAA waivers \cite{FederalAviationWaivers}, FAA policy orders \cite{FAA}, and the flight experience of our own team. 

We identified four high-level constructs of potential relevance for approving or denying a  flight request: \textit{drone characteristics, pilot history, operating environment, and proposed mission.} As a result of extensive discussions, we identified attributes associated with each construct to be evaluated for flight authorization decisions (e.g., pilot's flight hours). By systematically varying the levels of these attributes, we designed a diverse set of flight authorization request vignettes with the aim of gaining insights into how respondents would prioritize these factors when making decisions about flight authorization requests. The constructs, their varying levels, and the attributes used to determine the levels are shown in Table \ref{table:contructs}.

However, the combination of the levels of these constructs resulted in 81 ($3^{4}$) different vignettes, a number too large to be presented to each respondent, which is not uncommon in vignette-based studies \cite{atzmuller2010experimental}. In the interest of keeping the survey short and achieving higher completion rates, we selected nine vignettes that were broadly representative of the mix of factors. Table \ref{table:vignette_descriptions} shows the selected subset of vignettes based on the varying levels of the four constructs. Finally, we conducted a one-hour interview with a drone expert with over a decade of experience deploying sUAS for emergency response to solicit feedback on the quality and validity of our vignettes and the overall survey. This resulted in minimal design changes, specifically to the presentation of the information. The expert agreed with the general design of the vignettes but recommended we include more specific details about each mission. Figure \ref{fig:vignette} shows an example of a vignette used in the study. We provide both the preliminary and final design as artifacts on our supplementary website.\footnote{https://github.com/michaelchristopherhunter/DroneSurvey2023} The study was approved by Notre Dame University's Institutional Review Board (IRB) and was administered using the Qualtrics tool\footnote{https://www.qualtrics.com}.

\begin{table}[]
\caption{General outline for all nine vignettes in the study.}
\label{table:vignette_descriptions}
\begin{tabular}{c|llll }
\textbf{Vignette} & \textbf{Pilot}       & \textbf{Drone}   & \textbf{Conditions} & \textbf{Mission}   \\ \hline
V1       & Novice      & Average & Poor       & Difficult \\
V2       & Novice      & Poor    & Average    & Easy      \\
V3       & Novice      & Good    & Good       & Average   \\
V4       & Average     & Good    & Poor       & Easy      \\
V5       & Average     & Average & Average    & Average   \\
V6       & Average     & Poor    & Good       & Difficult \\
V7       & Experienced & Good    & Poor       & Average   \\
V8       & Experienced & Poor    & Average    & Difficult \\
V9       & Experienced & Average & Good       & Easy     
\end{tabular}
\end{table}

\subsubsection{Survey Structure}
The initial part of the survey collected demographic information about participants' backgrounds and drone flying experience, which we later used to explore associations between roles and perspectives. The participants were then presented with five randomly chosen vignettes, one at a time, from the nine different vignettes mentioned above. Each vignette prompted participants to make decisions regarding whether to authorize entry into the UTM system based on the presented circumstances. 

\begin{figure}[]
    \centering
    \includegraphics[width = \linewidth]{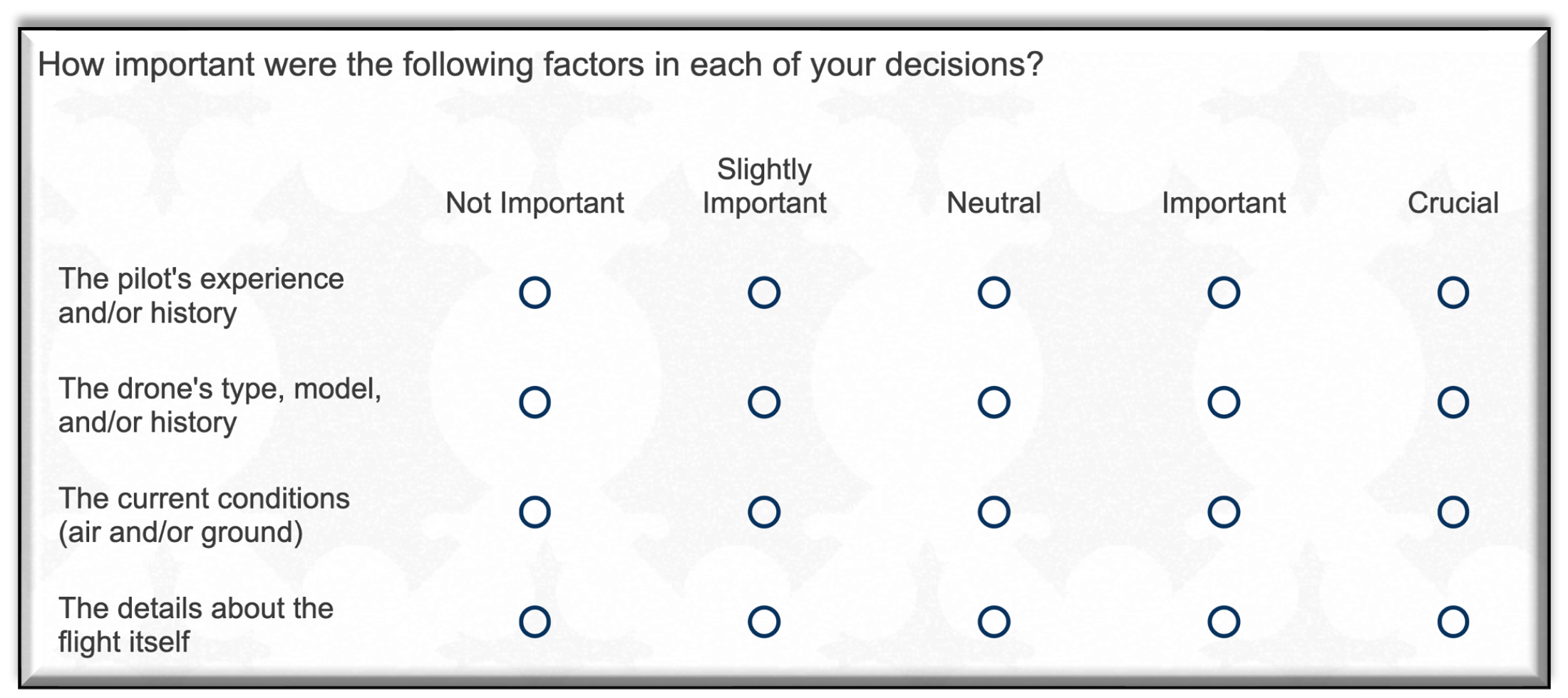}
    \caption{Participants rated factor importance for each vignette using a 5-point Likert scale.}
    \label{fig:questions}
\end{figure}

The participants were also prompted to rate each factor's importance in their decisions based on a 5-point Likert scale \cite{dawes2008data}, ranging from \textit{crucial} to \textit{not important} at all, as shown in Figure \ref{fig:questions}. Here, we \textit{only} assessed the importance of the high-level construct, not the low-level details, e.g., pilot history or the actual pilot attributes shown in Table \ref{table:contructs}. 

In the last section of the survey, we requested feedback regarding the overall significance of individual attributes in their decision-making on a sliding scale (0 to 100) \cite{roster2015exploring}. These included the following attributes of the planned flight:

\begin{itemize}[leftmargin=*]
  \renewcommand\labelitemi{--}
  \item The pilot's experience flying in similar weather conditions.
  \item The pilot's experience flying in similar environments (e.g., around buildings, congested airspace).
  \item  Number and severity of reported incidents the pilot has (e.g., crashes, rule violations).
  \item The time elapsed since 
  any pilot-related incidents (e.g., not following a rule).
  \item The drone's physical characteristics (e.g., size, speed) and physical features (e.g., onboard parachute).
  \item The drone's history, including its reported incidents, repairs, and/or regular maintenance.
  \item The drone's capabilities (e.g., detect and avoid, awareness of other aircraft).
  \item Current air traffic in the area.
  \item Current ground conditions in the area (e.g., traffic, obstacles).
  \item Current weather conditions in the area (e.g., wind, heat).
  \item The complexity of the planned flight (e.g., inspecting utility poles vs. flying in an open area).
  \item The characteristics of the planned flight (e.g., BVLOS, flying near people).
  \item The planning for the current flight (e.g., simulations, risk assessments).
\end{itemize}

This feedback on the significance of the individual attributes in their decision-making served two purposes: firstly, it gave us additional detailed insights into each participant's decision-making process, and secondly, it was utilized to understand how the interplay of factors can affect the decision. 
Lastly, we incorporated a qualitative section in the survey, soliciting participants' general perspectives on an automated flight authorization system in order to better understand societal concerns and community-related issues surrounding this system.

\subsection{Data Collection Process}
\noindent \textbf{Recruitment.} Our study targeted a diverse set of stakeholders, from drone hobbyists and enthusiasts to seasoned pilots and community members, with the aim of gathering feedback that was representative of the broad community affected by the automated flight authorization system.

We recruited participants in several different ways, as follows:

\begin{itemize}[leftmargin=*]
  \item \textit{Aviation day} at South Bend Airport. Several team members hosted a booth at a local aviation event with the primary aim of technology transfer of sUAS research deliverables. We used this opportunity to recruit people to the study by handing out invitation cards to any adult who visited the booth. Adults were primarily a mix of emergency responders, pilots (both airplane and drones), and community members.
  \item \textit{DroneResponders} is a non-profit organization of nearly 7000 members that unifies aerial first responders, emergency managers, and alike to enhance collaboration and training for more effective drone operations in public safety.
  \item Online discussion forums including \textit{Reddit}, \textit{Discord} groups, \textit{PX4} forums (discuss.px4.io) and \textit{Ardupilot} forums.
  \item Our own social media connections on \textit{LinkedIn} and \textit{Facebook}.
\end{itemize}

We refrained from offering participants any specific guidance apart from addressing basic technical and procedural queries.

\noindent \textbf{Data collection.} The study received a total of 96 responses. The responses encompassed a diverse group, including drone experts, hobbyists, and concerned community members, as reported in Table \ref{tab:demographics}. 
We reviewed each response individually to filter out any incomplete or invalid submissions. We included the responses that were more than 80\% completed as they were only missing final comments or the overall rating section. Furthermore, to enhance the integrity of our analysis, we excluded responses where participants consistently selected the same option for all questions and ratings. This left us with a total of 46 survey responses for use in the analysis. We recognize the inherent limitations of the size and sample of our responses. Nonetheless, it provides a good foundation to understand the valuable preliminary insights into diverse stakeholders' perspectives.

\begin{table}[t]
\centering
\setlength{\tabcolsep}{0pt} 
\caption{Self-reported characteristics of the respondents. Some participants selected multiple options.}
\resizebox{\columnwidth}{!}{
\begin{tabular}[width = \linewidth]{lc}
\toprule
\multirow{2}{*}{\textbf{Demographic}}           & \multirow{2}{*}{\shortstack[c] {\textbf{\# of responses that}  \\ \textbf{selected each demographic}}} \\
& \\
 \midrule
\textbf{Certified drone pilot (e.g., Part 107)} & 27   \\
\textbf{Drone Hobbyist (Certified)}             & 31   \\
\textbf{Emergency Responder}                  & 4   \\
\textbf{City Planner}                        & 2  \\
\textbf{Concerned Community Member}             & 9   \\
\hline
\end{tabular}}
\label{tab:demographics}
\end{table}

\subsection{Data Analysis}

We used Python and Excel to analyze responses from the quantitative part of the survey and employed the Scott-Knott ranking test \cite{6235961}. This significance test assigns ranks to the ratings, reflecting their relative importance, with higher ranks indicating greater importance. In our experiments, the ratings were ranked from 1st to 4th, with some factors sharing the same rank. For the open-ended qualitative part of the survey, where we ask participants to share societal concerns and issues about an automated flight authorization system, we used thematic analysis \cite{Clarke2014} with open coding \cite{miles1994qualitative,article1}. The first three authors independently analyzed the participants' comments and conducted open coding to organize the feedback into high-level categories. Following this, the authors met to review the codes and categorization to finalize a set of themes that best represented what we learned from our participants. We discuss these themes in the Results and Discussion sections.

\begin{figure*}[h]
    \centering
    \includegraphics[width = \linewidth]{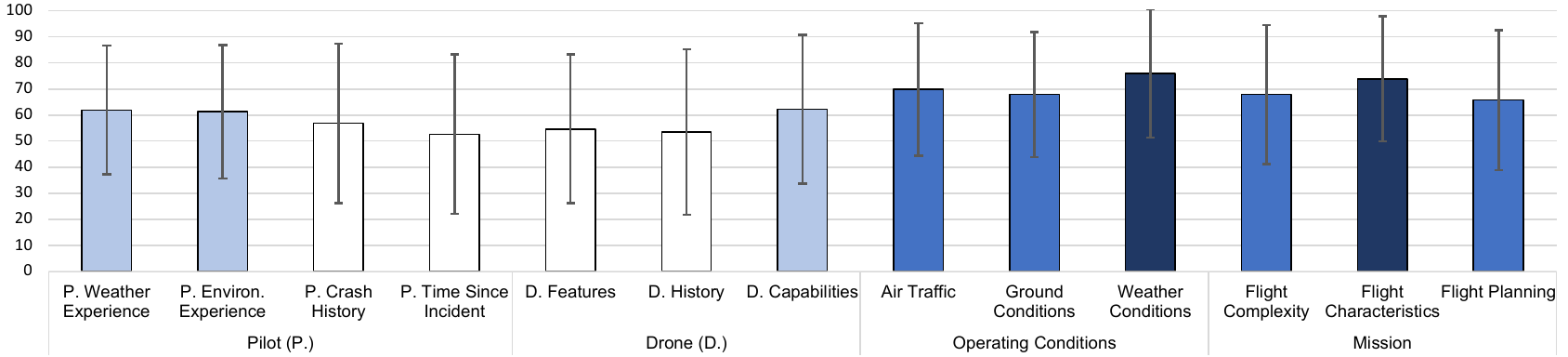}
    \caption{Overall mean ratings (0 - 100) on a slider scale for the general importance of each factor in the participants' decision-making. The error bars denote the standard deviation. The colors represent the ranked importance based on the Scott-Knott test \cite{6235961}. \colorbox{drk_blue}{\color{white}Dark}, \colorbox{med_blue}{lighter}, \colorbox{lt_blue}{light} and \colorbox{white}{white} represents the rank from first (most important) to fourth (least important).}
    \label{fig:ratings}
\end{figure*}
\section{Results}
\label{sec:results}

In this section, we present the key findings from our study, grouped by our research questions. We begin by analyzing the quantitative feedback to understand what factors stakeholders found important in their decisions \textbf{(RQ1)}. Then, we evaluate the qualitative portion of the study to explore participants' concerns regarding the design of the automated flight authorization system. \textbf{(RQ2)}.

\subsection{RQ1: Important factors in decision making}
\label{sec:RQ1}
We address RQ1, \textit{``What are the important factors that stakeholders report are needed to make fair, equitable, and safe decisions?''} by identifying three key factors.

\noindent--~\textbf{\textit{Operating environment \& mission complexity most critical to decisions. }}To determine which factors were considered most critical in the decision-making process, we examined the ratings that participants assigned to \textit{pilot, drone, operating conditions}, and \textit{mission elements} at the end of the survey, unrelated to any specific vignette. We ran the Scott-Knott significance test to rank the factors across all participants' feedback, confirming the statistical significance of our results.
The mean rating and ranking for each factor are shown in Figure \ref{fig:ratings}. Overall, the participants gave more weight to \textit{weather conditions} and \textit{mission characteristics} and less importance to \textit{pilot} and \textit{drone history}. More specifically, within pilot experience, participants placed greater importance on the pilot's prior experience in similar operating conditions (such as \textit{weather} and \textit{environment}) when compared to their \textit{crash history} and the \textit{time elapsed since any incidents}. The responses indicate that all four factors were considered relevant for their decision-making.

\noindent--~\textbf{\textit{Divergent views on fair decision.}} We also examined the responses for patterns associated with demographic characteristics. Demographics are reported in Table \ref{tab:demographics}. 

\begin{figure*}[h]
    \centering
    \includegraphics[width = \linewidth]{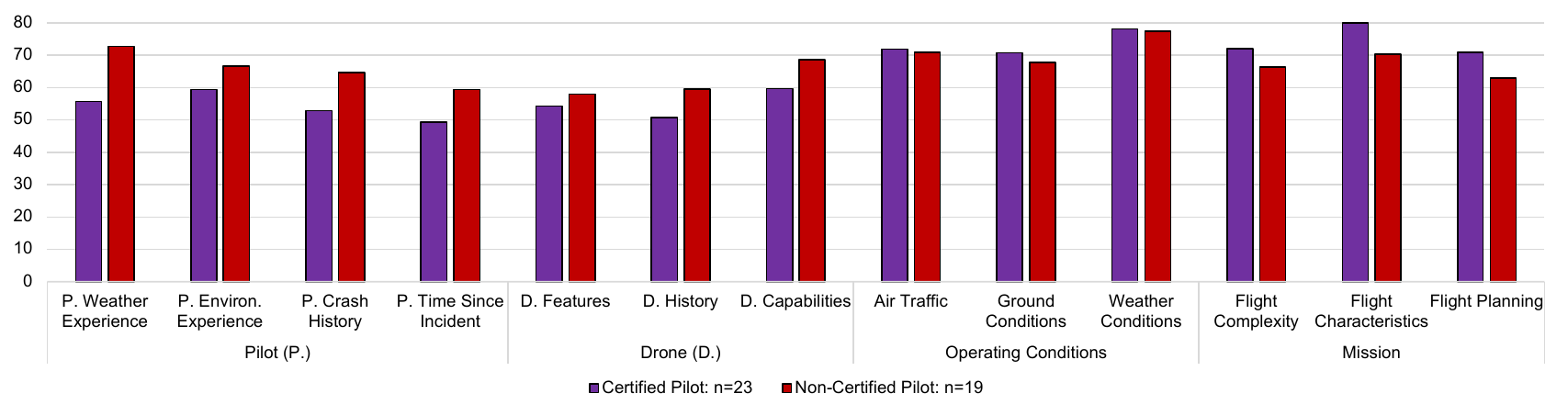}
    \caption{
    {Comparing overall mean factor importance ratings (0-100) on a sliding scale: Certified vs. Non-Certified Pilots.}}
    \label{fig:subgrouprating}
\end{figure*}

We observed a notable difference between the rankings of certified Part 107 pilots and non-certified pilots. As depicted in Figure \ref{fig:subgrouprating}, both groups ranked \textit{weather conditions} (from operating conditions) as very important. However, certified pilots ranked \textit{mission} parameters more highly than the non-certified pilots ranked them, while non-certified pilots ranked \textit{pilot experience} and \textit{drone history} more highly than the certified pilots ranked them. Furthermore, \textit{drone history} was considered least important by both groups.

Further investigation is needed to understand the reasons behind these differences; however, it is reasonable to expect certified pilots trained to follow safety procedures for each flight to provide more informed opinions. At the same time, it is important to understand the perspectives of a broader set of stakeholders. The implemented solution will need to have broad concurrence, and the rationales for specific design decisions will need to be broadly understood. 

\begin{figure*}[h]
  \begin{subfigure}[b]{0.58\textwidth}
    \includegraphics[width=\textwidth]{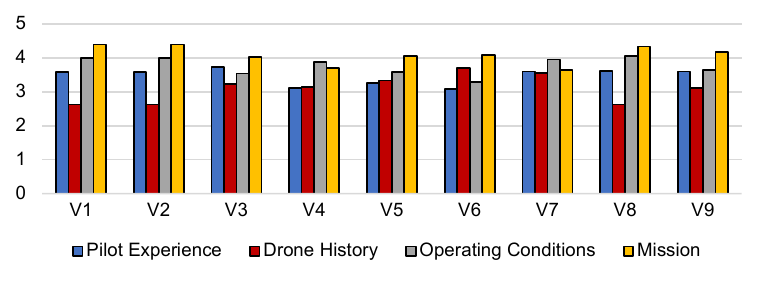}
    \caption{Average ratings of each factor across all nine vignettes.}
    \label{fig:1}
  \end{subfigure}
  \hfill
  \begin{subfigure}[b]{0.40\textwidth}
    \includegraphics[width=\textwidth]{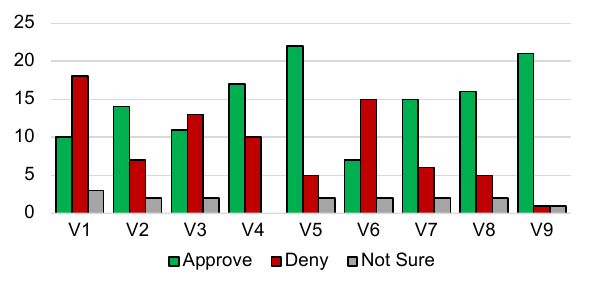}
    \caption{Overall number of approvals \& denials.}
    \label{fig:2}
  \end{subfigure}
  \caption{Survey Results}
  \label{fig:surveyresult}
\end{figure*}

\noindent--~\textbf{\textit{Context matters.}} One of the objectives of this study was to explore and demonstrate the complex nature of decision-making and illustrate that these decisions are far from being a straightforward "yes" or "no" as there are often competing factors. Even within the limited sample of scenarios presented in vignettes, we observed significant disparities in whether permission to fly should be granted or denied, as shown in Figure \ref{fig:surveyresult}(b). While several vignettes (V), such as V5, V9, and V8, achieved substantial consensus among respondents, others, such as V1 and V3, found little consensus. This highlights the challenges associated with the decision-making process and the importance of having clear and transparent criteria.

Consider V3, for instance, where a novice pilot with a prior incident requests authorization to fly a relatively uncomplicated mission (building inspection) under favorable conditions (see Figure \ref{fig:vignette}). The pilot was described as having limited experience and only one prior reported incident where they lost control and collided with a house whilst flying a \textit{different} drone. Respondents ranked pilot experience as the second most important factor for this particular vignette (second only to flight details). 11 respondents granted flight approval, 13 denied approval, and two were uncertain. The dilemma here is whether the pilot's previous incident with a \textit{different} drone is a negative factor in the decision. If yes, how many incidents are too many? Further analysis of the results revealed that approval and denial of permissions were not associated with pilot certification.  Among the 16 certified pilots, 10 denied and 5 approved, while among the 10 non-certified respondents, 6 approved and 3 denied. This illustrates the substantial degree of uncertainty surrounding these decisions, highlighting the magnitude of the challenge in establishing universally accepted and fair criteria for such determinations. 

\subsection{RQ2: Stakeholder concerns about automated UTM on-entry decisions}

\begin{figure}[h]
    \centering
    \includegraphics[width = \linewidth]{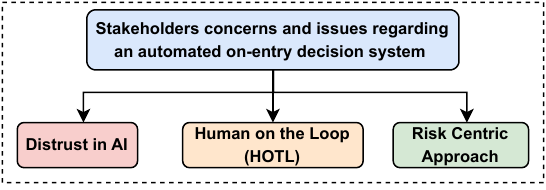}
    \caption{The themes identified from the responses.}
    \label{fig:themes}
\end{figure}

In this section, we answer RQ2, \textit{``What are the stakeholder concerns related to the automated flight authorization system?''} Three overarching themes emerged through our open-coding approach, as summarized in Figure \ref{fig:themes}. In the following discussion, we provide mappings to comments made by study participants (P1-P46) or left on social media sites (P*1-P*5) where we had posted survey invitations to the survey. 

\noindent--~\textbf{\textit{Distrust in AI.}} Some participants expressed a lack of trust in an AI-based authorization system. A primary concern that emerged was the necessity of implementing the system in the first place. The point was succinctly summed up by P13, who stated: "\textit{Don't use AI for this.}" Furthermore, a few participants raised concerns about the system's transparency and safety. For example, P40 said,\textit{``AI could easily misinterpret someone‚ $[$their$]$ intent and how $[$they$]$ word something".} This raises critical concerns regarding the structure and terminology used in describing a mission plan when issuing a flight authorization request, in order to minimize the potential for AI to misinterpret inputs, such as mission characteristics, and, even more significantly, to reduce vulnerability to malicious attacks — a well-documented issue in literature \cite{qiu2019review}. Another participant noted the importance of transparency, stating that  \textit{``I'm not sure I'm keen on an algorithm deciding if I can fly my drone, and I feel there needs to be a bit more transparency in this survey and project." (P*5)}. This comment reflected not only the high-level nature of our vignettes but also highlighted a clear need for transparency regarding how authorization decisions are made.


\noindent--~\textbf{\textit{Human on the loop (HOTL).}} In contrast to the respondents who 
rejected an AI approach based on lack of trust, others recognized the potential value of AI-supported decision-making while advocating for a "Human-on-the-Loop" (HOTL) \cite{10.1145/3468264.3468534} approach. In a HOTL system, the 
software can autonomously make decisions, but a human operator actively monitors the system and retains the ability to intervene when needed. Many participants expressed that human oversight, evaluation, or intervention was needed as the combined effect of different factors can make the decision very complex. For example, \textit{``This should never be left up to AI. There should always be a person involved when it comes to approving more complex missions. The human element cannot be taken out of the equation."} (P33)

An automated system must balance safety concerns with efficiently approving low-risk drone flights. However, we must determine the conditions under which human review is needed. A speedy but rigorous and systematic safety analysis is needed for all automated decisions. One instance where this can work is in emergency scenarios; as one participant pointed out, \textit{``How would AI be able to determine the exigency or severity of an incident?"} (P*3) 

\noindent--~\textbf{\textit{Risk-centric approach to decisions.}} One prominent theme that emerged through the analysis of the open-ended questions was that some participants felt that the decision should be entirely based on risk assessment and mitigation. While acknowledging the significance of factors identified in the survey, such as the pilot's experience, they prioritized the safety of people in the vicinity of drone missions. As P3 put it: \textit{``Decisions should be primarily based on safety concerns, more specifically on whether there is a possible threat to the people in the vicinity of the mission. The drone should be capable of flying under current weather conditions and should have many layers of safety that mitigate possible damage to property or people (parachute, detect and avoid, etc.). The pilot's experience, while important, is not a major cause of concern (if the drone navigation system is not overly complex)."}  This perspective clearly reflects the FAA's general safety-oriented approach to approvals \cite{policyorder}.

However, a few respondents contended that in most cases, the potential risks to crewed aviation and people on the ground were negligible or close to zero, making a compelling case for approving these flights. For instance, one respondent (P*4) said, \textit{``The entire manned aviation system is based around minimizing fatalities. By that metric, sUAS has a perfect record, yet the FAA restricts..."}

These divergent viewpoints reflect a complex balance between risk management and airspace accessibility in regard to sUAS flight authorizations. While one group calls for a robust risk-based approach, prioritizing safety, the other suggests that the current framework may, at times, be overly cautious given the inherently low risk posed by most drones. These differences in perspective underscore the nuanced considerations required for the development of an effective and widely accepted automated authorization system.

\section{Lessons Learned and Future Work}
\label{sec:discussion}

In this section, we present a more in-depth discussion of our results and the lessons we have learned, and we propose future work. 

\subsection {Lessons Learned}
\noindent \textbf{
Reluctance to accept fully autonomous decision-making.} Among the respondents, there was notable hesitancy about adopting an autonomous or AI approach for sUAS flight authorizations; however, it is common for AI systems to face initial skepticism \cite{longoni2019resistance,hengstler2016applied}. The complexity of the decision-making process in the context of sUAS flight authorizations, coupled with concerns about the reliability and safety of AI-driven decision systems \cite{insights}, may impact the broad acceptance of such a system. The survey also revealed concerns about transparency--that the criteria for admitting or denying entry to airspace might be opaque, especially if AI were involved. This reticence is amplified by the expressed concerns that AI might make incorrect decisions, misinterpret data, or lack the ability to adapt to unforeseen circumstances, especially in the dynamic and often unpredictable environment of sUAS operations. 

Based on survey comments, approaches that include human intervention or supervision seem likely to improve acceptance and trust. However, the need for scalability and rapid decisions will complicate the tradeoffs. The alternative of slow, manual decisions on all but very simple entry requests cannot scale to the number of sUAS that must be served. 

\vspace{3mm}
\noindent \textbf{Equitable access through broad community engagement.} Some participants raised valid concerns about airspace regulation, prompting questions about equitable access. There are concerns that the rapid expansion of drone operations, particularly for commercial purposes, could impose restrictions on hobbyists, limiting their access to the airspace \cite{GROTE2022102218}. The question arose as to whether this would be fair. Furthermore, potential regulatory requirements for costly specialized equipment could be prohibitively expensive for non-commercial users \cite{decker2022economic}. One cited example was the recent FAA requirement for drones flying outside of FAA-recognized identification areas to be equipped with remote ID capabilities, creating a new barrier. Such concerns, if not addressed in the new system, may foster a sense of exclusion, with negative consequences for wider acceptance \cite{GROTE2022102218}. 
Perhaps unsurprisingly, stakeholders did not always agree on whether a flight request should be approved or not. While experts play a crucial role in shaping future policy decisions, our findings underscore the necessity of soliciting input from the diverse community of potentially impacted stakeholders to understand subgroups' concerns, such as drone hobbyists.

\vspace{3mm}

\noindent \textbf{Safety awareness.}  The biggest difference between certified pilots and non-certified respondents was that the former approved fewer scenarios than the latter. 
We speculate that this is because the FAA regulation (Part 107)  that certified  pilots follow trains them to perform a mental hazard analysis and risk mitigation.  It includes safety requirements such as 
    "No person may operate a civil small unmanned aircraft system unless it is in a condition for safe operation" and   
"No person may: 
(a) Operate a small unmanned aircraft system in a careless or reckless manner so as to endanger the life or property of another." 
Certified pilots, familiar with regulations and factors ensuring airspace safety, were more risk averse than non-certified participants in approving scenarios. It may be, for example, that the limited information about the scenarios caused the certified pilots to be cautious about approving them. Perhaps they would approve more scenarios if given more detailed information, including additional factors.  

\subsection {Future Work}
The study results indicated some actionable stakeholder concerns that inform the design of an automated flight authorization system. Perhaps unsurprisingly, even among certified pilots, there was a lack of consensus about whether a drone could safely enter shared airspace in certain scenarios. Many participants were also reluctant to rely on automated software to make this decision. To address these concerns, we plan to explore ways to engage humans in the decision-making process while automating the first-line decisions by including humans in appeals and/or more complex cases. 
Furthermore, rather than granting or denying authorization, a third option might be added where, in some cases, the sUAS receives conditional authorization with associated flight constraints.

A typical flight authorization request provides detailed information about the proposed flight; however, our study used a simplified abstraction of a request in order to keep the survey short and maximize the completion rate. In the future, we plan to investigate more diverse scenarios and conduct focus studies and semi-structured interviews with the larger community to gain a deeper understanding of their concerns and elicit their input on automated flight authorizations. Future work also will include studies based on detailed authorization requests and their evaluations.

Some participants suggested adding more details to the authorization requests, such as the actual number of flying hours and all-up weight, for a more comprehensive evaluation. We hope to investigate how best to incorporate all the information. Additionally, while we had both non-pilot community members and city planners in this initial survey, we plan to broaden the scope of our study by including a wider spectrum of people who are stakeholders affected by airspace decisions (homeowners, school administrators, small local business owners), but are not drone pilots. 

Certain requirements for the automated flight authorization system will inevitably evolve over time. One aspect of this may be an evolution from using heuristics to rule-based automation to ML/AI-enabled decision support. Whatever underlying approach is used, it will be called upon to be transparent in its decision-making, to compute fair and equitable decisions, and to contribute to safe airspaces. This study is a first step in our larger project, with survey findings guiding the structure of future focus groups and subsequent steps.

\section{Threats to Validity}
\label{sec:threats}

\textbf{Internal validity.} Limited participant numbers and demographic diversity may impact our results. Additionally, our survey was limited to US participants and the FAA UTM system  \cite{10.1145/3411764.3445488}. Finally, the chosen vignettes may not encompass all identified factors, potentially omitting valuable scenarios However, this initial exploratory study will be succeeded by a more comprehensive one, featuring detailed flight requests, open-ended interviews, and a larger, more diverse participant pool.

\textbf{External validity.} 
To minimize external threats we adopted a comprehensive outreach strategy when inviting participants, resulting in a relatively diverse pool that encompassed not only drone experts in the industry but also hobbyists. However, the limited size of the study means that results may not represent the full spectrum of opinions. We will address this through our planned study. 

\textbf{Recall bias.} Responses may be prone to memory bias, with participants relying on initial recollections and potentially overlooking relevant details. Additionally, framing bias, influenced by the presentation style and random sequencing of vignettes \cite{framing,aouad2021display}, could impact responses. While inherent to random sequencing, a larger follow-up study might help mitigate this bias.

\textbf{Construct threats.} To mitigate construct threats, we consulted an expert in deploying sUAS for emergency response, before the study. Moreover, we used open coding \cite{miles1994qualitative} and best practices \cite{10.1145/3510458.3513008}, including multiple discussion sessions, to avoid misinterpreting qualitative data. However, a potential construct threat arises from a low response rate for the qualitative part of the survey.


\section{Related Work}
\label{sec:related}

\noindent \textbf{Operational Protocols in Shared Airspace.} A significant body of work has proposed \cite{allouch2021utm,9473838,2021,doi:10.2514/6.2017-3273} and evaluated \cite{alarcon2020procedures, xiangmin2020survey} protocols to handle high-capacity situations for sUAS in shared airspace. This work spans autonomous flight path planning (conflict resolution, detect and avoid systems) \cite{9473838}, risk assessment frameworks \cite{doi:10.2514/6.2017-3273}, and UTM software architectures with automated threat management \cite{2021}. These studies mainly focus on automation in UTM \textit{after} a drone has entered the airspace, which is not the focus of our work.


\noindent \textbf{Community Perspectives on Shared Airspace.} There have been multiple studies conducted to elicit feedback from the general aviation community on the integration of drones in shared airspace with other aircraft and users \cite{GROTE2022102218}. In the context of the European equivalent of UTM (U-Space), \citeauthor{barrado2020u} \cite{barrado2020u} highlighted the essential services necessary to support such an ecosystem. These services encompass both pre- (such as drone registration and weather information) and post-flight services (e.g., strategic conflict resolution), as well as in-flight services like electronic identification, position reporting, etc. In another study \cite{merkert2020managing}, an examination of drone usage revealed a shift from historical concerns like privacy and security to pressing operational issues, notably interactions with other users. It underscored the need for regulations and restricting drone access, motivating our study. A follow-up survey \cite{merkert2021will} revealed that drone operators were willing to pay for access to restricted ecosystems, similar to road tolls. Finally, \citeauthor{decker2022economic} \cite{decker2022economic} highlighted key factors for a UTM economic framework, including safe airspace access and data sharing.  

\citeauthor{bauranov2021designing} \cite{bauranov2021designing} examined various global airspace concepts falling under the UTM umbrella. They noted that these concepts primarily emphasized safety and capacity optimization, often overlooking social factors like public acceptance issues such as noise, visual disruption, and privacy concerns. Our study is an initial attempt to get feedback from the community to develop a broad consensus on automated flight authorization systems. Finally, \citeauthor{GROTE2022102218} \cite{GROTE2022102218} studied the views of the General Aviation community in the UK regarding issues and concerns about the UTM concept. This work is the closest to our study; however, it primarily focused on summarising concerns of the community about potential issues within the airspace. Instead, our work seeks the community's feedback on an automated flight authorization system.

\section{Conclusion}
\label{sec:conclusion}

As increasing drone adoption accelerates the need for automated low-altitude flight authorization software systems, ensuring that these systems are fair and equitable toward wide public acceptance is important. In this paper, we report results from an initial study that investigated the perspectives of the larger sUAS community on such systems. Using a vignette-based approach, we explored key factors influencing stakeholders' perceptions of fairness and safety in the flight authorization decision-making process. We found that there were divergent views on what factors are important for making such a decision, potentially impacting equitable access. Additionally, results revealed concerns about automated solutions, especially if AI-assisted. Our findings underscore the need for future research that engages a diverse and inclusive spectrum of participants while carefully considering safety concerns related to more detailed flight authorization requests.

\begin{acks}
This work was funded by grant 80NSSC23M0058 from 
NASA. We thank the anonymous reviewers for useful comments. 
\end{acks}


\balance
\bibliographystyle{ACM-Reference-Format}
\bibliography{sample-base}



\end{document}